# The significance of user-defined identifiers in Java source code authorship identification

Georgia Frantzeskou[#], Stephen G. MacDonell[*], Efstathios Stamatatos[#],
Stelios Georgiou[¶], Stefanos Gritzalis[#]

[#] *Dept of Info. and Comm. Systems Engineering*
*University of the Aegean,*
*Samos, 83200, Greece*
*Email: (gfran, stamatatos, sgritz)@aegean.gr*

[*] *SERL, School of Comp. and Math. Sciences*
*Auckland University of Technology,*
*Private Bag 92006, Auckland 1142, New Zealand*
*Email: smacdone@aut.ac.nz*

[¶] *Department of Statistics and Actuarial - Financial Mathematics*
*University of the Aegean,*
*Samos, 83200, Greece*
*Email: stgeorgiou@aegean.gr*

**Abstract**

*When writing source code, programmers have varying levels of freedom when it comes to the creation and use of identifiers. Do they habitually use the same identifiers, names that are different to those used by others? Is it then possible to tell who the author of a piece of code is by examining these identifiers? If so, can we use the presence or absence of identifiers to assist in correctly classifying programs to authors? Is it possible to hide the provenance of programs by identifier renaming? In this study, we assess the importance of three types of identifiers in source code author classification for two different Java program data sets. We do this through a sequence of experiments in which we disguise one type of identifier at a time. These experiments are performed using as a tool the Source Code Author Profiles (SCAP) method. The results show that, although identifiers when examined as a whole do not seem to reflect program authorship for these data sets, when examined separately there is evidence that class names do signal the author of the program. In contrast, simple variables and method names used in Java programs do not appear to reflect program authorship. On the contrary, our analysis suggests that such identifiers are so common as to mask authorship. We believe that these results have applicability in relation to the robustness of code plagiarism analysis and that the underlying methods could be valuable in cases of litigation arising from disputes over program authorship.*

## 1. INTRODUCTION

Although programming languages are generally more formal and restrictive in their form and composition than spoken or written languages, program authors are still afforded a large degree of flexibility when composing source code [1]. Each programming language has a vocabulary of keywords, reserved words and operators, from which program authors select appropriate terms during the programming process [2]. In addition, source code contains vocabularies of identifiers (names of variables, procedures, functions, modules, labels and the like) created by programmers. Identifier naming can be influenced by many things – the application domain, agreed coding styles, organisational guidelines, or an educator's advice in the case of code written by students. Although the importance of meaningful identifiers has been stressed for many years by educators, Sneed observed that "in many legacy systems, procedures and data are named arbitrarily…. programmers often choose to name procedures after their girlfriends or favourite sportsmen [*sic*]" [3]. Thus while coding styles and the like exist, the degree to which these actually influence identifier naming – and the consequent impact on authorship analysis – is unknown. It may be that programmers create their identifiers in a systematic or consistent way so that any resulting program reflects its author. With this in mind, the aim of this paper is to assess the impact that identifiers have on the accuracy of Java source code authorship attribution, using two sets of Java programs. In other words, the questions we address here are: Do Java identifiers contribute to correct authorship identification? Is it possible to hide the provenance of some Java program by identifier renaming?

These questions are very important whenever a need for evidence arises in regard to source code authorship [4], such as in formal dispute proceedings. For example, we may be able to assert with evidence that Java programmer A is the author of a disputed program because the class variables and/or method names used closely resemble those used elsewhere by programmer A – and because they do *not* resemble those used by one or more other

Java programmers. While some of our recent work [4] has indicated that Java identifiers defined by the programmer do not influence classification accuracy, and in fact in some cases accuracy might be improved by 'neutralizing' these names, that study examined all user-defined identifiers together. In this paper we check whether this conclusion holds when we examine each type of programmer-defined identifier separately.

In conducting our analysis, we use the SCAP approach as a tool for assessing the significance of each type of Java identifier. The use of Source Code Author Profiles (SCAP) represents a new approach to source code authorship identification and classification that is both highly effective [5, 6] and language-independent, since it is based on low-level non-metric information. In SCAP, byte-level n-grams are utilized to assess the form and nature of code against known author profiles (described in more detail in Section 3). We have chosen to use n-grams in this method as they are more flexible and expressive in comparison to fixed lists of tokens used in previous studies [6, 7]. Kothari at al. [7] for instance, have demonstrated in a recent study using two large scale Java data sets, that n-grams significantly outperformed style based metrics. A further advantage of n-grams is that they can be extracted without the need to construct special parsing or mining tools.

Three different types of Java identifiers are considered here: class, method and simple identifiers (defined in Section 4). A number of experiments have been performed in order to assess the impact that each type of identifier has on source code authorship attribution. We first assess classification accuracy using the complete source programs, creating an initial performance benchmark. We then measure the contribution of each identifier type in authorship identification by running a sequence of experiments, each time disguising a certain type of identifier in the source code. We are therefore able to measure the effect that 'neutralization' of each type of identifier has on the accuracy of the classification – the difference in each case effectively indicates the relative significance of this type of identifier to authorship identification. The experiments have been performed using two different data sets for which the authors of the programs are known. One data set comprises open source programs and the second is made up of programs written by students during an introductory Java course.

The remainder of this paper is organized as follows. Section 2 addresses prior studies related to the use and influence of program identifiers. A detailed description of the SCAP approach is provided in Section 3. Section 4 describes the types of Java user-defined identifiers that are thought to influence source code authorship identification. Section 5 describes the two Java data sets used: the Open Source and the Student data set. Section 6 details all the experiments performed on the two data sets in order to examine which (if any) identifiers contribute to authorship identification, and to what degree. Finally, Section 7 summarizes the outcomes of the empirical work, reports the conclusions drawn from this study and proposes future research directions.

## 2. RELATED WORK ON IDENTIFIERS IN PROGRAM CODE

The principal research interests associated with the use of identifiers in and of source code have tended to relate to program comprehension and concept location. For instance, the study of Anquetil and Lethbridge [8] dealt with the analysis of file names, with the purpose of clustering those names that reflect common concepts. In particular, the problem of automatically building an abbreviation dictionary to segment file names into their component abbreviations was discussed. It was constructed by exploiting alternative sources, including an English dictionary and the set of n-grams shared among file names, comments and function identifiers. While their study does not address authorship as such it is of relevance in its use of n-grams, a method not commonly employed in source code authorship analysis but that we use here in SCAP. In Merlo et al.'s study [9], sequences of words taken from comments or identifiers were classified against a concept tree provided by a human expert, by means of an artificial neural network. Merlo et al. state that "Identifiers and comments are important sources of information". The selection and use of meaningful identifiers is elsewhere considered useful in terms of enhancing code or design traceability [10, 11] and as a basis for considering design evolution [10].

Takang et al. [12] sought to determine the contribution of identifiers and comments to program comprehension. They assessed comprehension performance for code that utilized abbreviated identifiers as compared to full-word identifiers, and that contained either uncommented or commented code. The results showed that programs that contained full word identifiers were more understandable than those with abbreviated identifiers; however, Takang et al. were unable to arrive at any definitive conclusions regarding the use of potentially more informative abbreviations, created from the first two letters of each word. In addition the results showed that commented programs were more understandable than non-commented programs. While this is an interesting and potentially useful result, its general applicability to other types of programs in other domains is unknown since only a single program was used in the analysis, and its goal of program comprehension is different to one of authorship classification. That said, it is relevant to the work reported in this paper in terms of its use of abbreviated identifiers, in that the two-letter abbreviations used by Takang et al. could be considered as equating to initial bigrams.

Caprile and Tonella [13] analyzed function identifiers by considering their lexical, syntactical, and semantic structure. For this purpose they applied a segmentation technique to function names in order to build a dictionary of words used as components. By following a grammar they anticipated improvements in the readability, understandability and, more generally, the maintainability of a program. In a study that followed they presented an approach for restructuring program identifiers aimed at improving their meaningfulness [14], by exploiting a standard lexicon for word composition and a standard syntax for their arrangement. The approach has been implemented in two applications. In both cases, names could be migrated to a more meaningful standard form, thus increasing their self-documenting ability. Such manipulation could clearly have an impact on authorship analysis – the extent to which such standardisation practices are used, however, is not clear at this time.

Deißenbock and Pizka [15] created a formal model of concepts and name spaces in code. It was asserted that this model would provide a solid foundation for the definition of precise rules for concise and consistent naming in source code development. A tool was created which incrementally built and maintained a complete identifier dictionary while the associated system was being developed. The authors contended that this identifier dictionary would aid in consistent naming, and improve productivity of programmers by proposing suitable names depending on the current context. Deißenbock and Pizka argued that naming conventions are needed to enforce consistency and to provide guidelines about the mechanics of turning a concept into an appropriate name. With such guidelines, names should contain enough information for an engineer to comprehend the precise concept. The work of Deißenbock and Pizka has been extended by the work of Lawrie et al [16]. They identified limitations in the Deißenbock and Pizka work and proposed an augmented variable naming approach that does not require any additional information (e.g. a mapping). Lawrie et al. [17] have also conducted a study on program comprehension based solely on identifier naming. The study involved over 100 programmers who were asked to describe twelve different functions. The functions used three different "levels" of identifiers: single letters, abbreviations, and full words. The results from this study showed that better comprehension is achieved when full word identifiers are used rather than single letter identifiers. It also showed, however, that in many cases abbreviations are as useful as full word identifiers. In terms of the work undertaken here, the use of abbreviations could potentially confound the ability to identify code authors, in that (i) they may be used generally by more than one author, and (ii) they reduce the n-gram space that can be traversed in order to build programmer profiles.

Kuhn et al [18] presented an approach to retrieve topics evident in source code by exploiting linguistic information found in identifiers and comments. They claimed that "Source code bears the semantics of an application in the names of identifiers and comments". Their approach, called semantic clustering, is based on Latent Semantic Indexing and clustering to group source documents that use similar vocabulary. The case studies examined give evidence that this approach provides a useful first impression of an unfamiliar system, and that it can reveal valuable developer knowledge. Their initial hypothesis that semantic clustering would reveal a system's domain semantics did not prove to be supported since their experiments showed that most linguistic topics were related to application concepts or architectural components rather than the underlying domain. This indirectly suggests that identifiers may be program-specific and reflective of their composer, a useful premise for authorship attribution based on such constructs.

In our own study [4] we examined the significance of high-level programming features in terms of their influence on source code author classification. We used programs written in two languages so as to consider two different programming styles: Java, which uses objects, and Common Lisp, which uses a functional/imperative programming style. A range of features were considered and the importance of each in determining authorship was measured through a sequence of experiments in which we disguised one feature at a time. The results (although not tested for statistical significance) showed that, for these programs, comments, layout features and package-related naming influenced classification accuracy whereas user-defined naming, an obvious programmer related feature, did not appear to influence accuracy.

The majority of the studies considered above analyzed identifier naming in relation to concepts described within a piece of code, and the impact that this might have on understanding. While this is peripherally relevant, little work to date has specifically assessed the role of identifiers in reflecting and determining program authorship. The studies above do emphasize the importance of identifier meaningfulness in enhancing understanding. However, since virtually every programming language allows programmers to use almost arbitrary sequences of characters as identifiers, in principle any given programmer has the freedom to use as an identifier name any sequence of characters he/she chooses, potentially resulting in meaningless or even misleading – but perhaps author-specific – naming [15]. The questions we therefore address here are: Do Java identifiers contribute to correct authorship identification? Is it possible to hide the provenance of some Java program by identifier renaming? The following sections describe our efforts to answer these questions.

## 3. THE SCAP APPROACH

This section describes in detail the tool used to assess the significance of identifiers (and other features) in our experiments to date addressing authorship attribution [4,5,6]. It is called the Source Code Author Profiles (SCAP) approach and is an extension of a method that has been successfully applied to text authorship identification by Keselj et al. [19]. The approach is based on the extraction and analysis of byte-level n-grams, present in our case in program source code. An n-gram is an n-contiguous sequence that can be defined at the byte, character, or word level. For example, the byte level 3-grams (or trigrams) extracted from 'The first' (where the character _ indicates a space) are: The, he_, e_f, _fi, fir, irs, rst. Byte, character and word n-grams have been used in a variety of applications such as automatic text categorization [20], authorship attribution [19], optical character recognition [21], string matching [22].

The SCAP procedure is explained in the following steps and is illustrated in Figure 1. The bolded numbers shown in Figure 1 indicate the corresponding step in the description that follows. In essence, it calculates the most likely author of a given piece of code for different values of n-gram size n and profile length L.

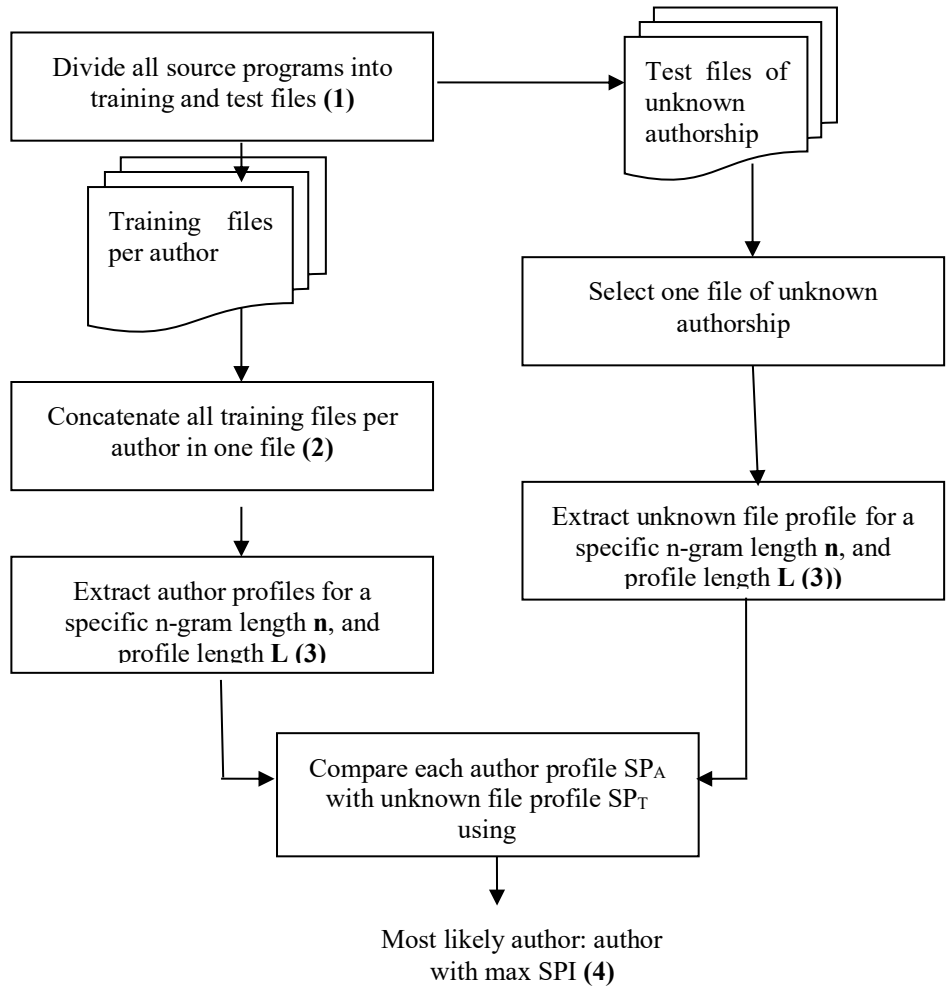

**Figure 1.** Estimation of most likely author of an unknown source code sample using the SCAP approach

1. Divide the known source code programs for each author into training and testing data sets. (We have tended to split the sets evenly; however, the particular allocation of proportions may vary depending on the size of the total sample, the number of programs per author, the desire for greater confidence in analysis and so on.)

2. Concatenate all the programs in each author's training set into one file. Leave the testing data programs in their own files.

3. For each author training and testing file, get the corresponding profile:

   - Extract the n-grams at the byte-level, including all non-printing characters. That is, all characters, including spaces, tabs, and new line characters, are included in the extraction of the n-grams. In our analyses, Keselj's [23] Perl package Text::N-grams has been used to produce n-gram tables for each file or set of files required.

   - Sort the extracted n-grams by frequency, in descending order, so that the most frequently-occurring n-grams are listed first. The n-grams extracted from the training file correspond to the author profile, which will have varying lengths depending on the length (in terms of characters) of the programming data and the value of n (n-gram length). The profile created for each author is called the Simplified Profile (SP).

   - Keep the L most frequent n-grams $\{x_1, x_2,…,x_L\}$. The actual frequency is not used mathematically except for ranking the n-grams.

4. For each test file, compare its profile ($SP_T$) to that of each author ($SP_A$) using the Simplified Profile Intersection (SPI) measure:

   - Select a specific n-gram length, such as trigram (for the experiments in this paper, we used a range of lengths, trigrams up to 10-grams). Select a specific profile length L, at which to cut off the author profile, smaller than the maximum author profile length.

   - For each pair of test and known author profiles, calculate the SPI measure. Letting $SP_A$ and $SP_T$ be the simplified profiles of one known author A and the test or disputed program T, respectively, then the similarity distance SPI is given by the size of the intersection of the two profiles:

   $$|SP_A \cap SP_T|$$

   In other words, the similarity measure we propose is the number of common n-grams in the profiles of the test case T and the author A.

   - Classify the test program to the author whose profile at the specified length L has the highest number of common n-grams with the test program profile at the specified length. In other words, the test program is attributed to the author with whom the largest amount of intersection is achieved.

We have developed a number of Perl scripts in order to create the sets of n-gram tables for the different values of n (n-gram length), L (profile length) and for the classification of each test program file to the author with the smallest distance (i.e. the greatest overlap of n-grams). By shifting the n-gram length n and the profile length L (the number of n-gram types included in the SPI), we can test how accurate the method is under different n, L combinations.

## 4. JAVA IDENTIFIERS AND SOURCE CODE AUTHORSHIP IDENTIFICATION

As stated above, computer programs are written according to strict grammatical rules (context free and regular grammars) [24] and are (typically) constructed using one language. Each programming language has its own vocabulary of keywords, reserved words and operators, from which program authors select appropriate terms depending on the functionality required [25]. In addition, program authors create vocabularies of numbers and vocabularies of identifiers (names of variables, methods, constants, modules, labels and the like). These are, in general, not language dependent. When using Java to construct a program the author must 'create' some words, such as class names and method names [26]. Our aim in this study is to measure the significance of programmer-defined Java identifiers in relation to source code authorship identification. We have considered these identifiers in terms of four main categories:

1. <u>Simple Identifiers:</u> This category includes all variables defined within the program using one of the primitive data types defined in Java, e.g. int, char, byte, long, boolean, int[] and long[]. Common examples of such variables might be year with type int or flag with type boolean.

2. <u>Class Identifiers:</u> This category includes all variables that represent objects in the program and all names of classes defined within the program. For example, name with type String or ex with type Exception.

3. <u>Method Identifiers:</u> This category includes all method names that are defined within the program. Some examples are getInteger, drive and getYear.

4. <u>All Identifiers:</u> All user-defined identifiers, effectively the aggregation of those elements in the three categories listed.

Other source code terms, such as String and System.out.println, are not created by the person who writes the piece of code but are drawn from the Java API and are simply selected for use by the programmer. These names are not part of our current study, although it is acknowledged that they might influence classification accuracy of Java programs [4].

## 5. DATA SETS ANALYSED

In order to assess the contribution of user- (or programmer-)defined identifiers to authorship identification of Java programs, two different data sets have been considered here. The first data set included open source (OS) programs written by eight different authors as found in the freshmeat.net web site. The programs were allocated to approximately equally-sized training and test sets. In order to make the allocation and analysis 'domain independent' all programs from each author that were placed in the training set were from a different project from the programs placed in the test set. Hence, we had programs from sixteen different projects, two projects for each author. Consequently, since they were from different projects the programs in each set for each author did not share inherently common characteristics. The total number of programs in this data set was 35. Eighteen programs were allocated to the training set and 17 to the test set. This data set is from this point referred to as the OSJava data set.

The second data set comprised programs written during an introductory Java course. These programs were written by 8 different programmers, making up a sample of 54 programs in total. The data set was split into quasi equally-sized training and test sets. As these programs were student assignments from an introductory programming course there is a high degree of likelihood that the identifiers used will have been influenced by the guidance of the instructor, and perhaps by that found in course texts. More significantly, we know that some of the source code samples had been plagiarized. In addition, most of the program samples are from the same domain (i.e. sorting algorithms, binary search). All these facts imply that the programs in this data set potentially share several common characteristics. This data set is from this point referred to as the StudentJava data set.

In order to establish suitable performance benchmarks for the two data sets we removed all comments from the OSJava and StudentJava programs and ran a first SCAP authorship analysis experiment with all identifiers intact. The comments were removed because our aim in the tests was to consider the degree to which identifiers in the source code contributed to authorship identification, without the 'influence' of comments. The profile sizes used ranged from 2000 up to 8000 n-grams (in steps of 1000), whereas the n-gram sizes used were from 3-grams up to 10-grams. (These parameter sizes apply to all subsequent experiments presented in this paper. They have been selected because they have been shown empirically to be the most effective sizes in authorship attribution [4, 7]. The percentage of correctly classified test programs was recorded for each combination of profile and n-gram size, producing 56 (7 x 8) results per experiment per data set.

## 6. SIGNIFICANCE OF IDENTIFIERS

Focusing on each of the four categories of identifiers that might influence authorship identification in turn, as described in Section 4, a set of experiments was performed on both the OSJava and StudentJava program sets in order to measure the contribution of each type of identifier to accurate classification. Hence, for each category of identifier we report the results obtained from both the OSJava and the StudentJava data sets after disguising the relevant identifiers and then analyzing and classifying the programs using the SCAP approach. If a category of identifiers is positively influential in reflecting program authorship then we would expect that classification performance would deteriorate if those identifiers were 'hidden' from the analysis (shown as 'Worse than OSJava/StudentJava' in Tables 1 and 2). On the other hand, if a certain group of identifiers was not influenced by their authors then we would likely see the same levels of performance achieved in the allocation of test programs to authors as that achieved in the benchmark tests (i.e. 'Same as OSJava/StudentJava').

In order to assess the statistical significance of the results we obtained, we performed one-tailed t-tests to compare the mean accuracy achieved in the benchmark OSJava and StudentJava tests and the mean accuracy obtained after disguising each identifier type. The significance level used was 5% for all experiments. Note that we first checked whether our data were normally distributed and it was found that they were approximately normal. To add further reassurance, however, we also analysed our results using the nonparametric Wilcoxon test and arrived at outcomes consistent with those obtained from the t-tests.

Table 1 shows a summary of the classification accuracy results achieved for the four OS data set experiments across the various combinations of SCAP profile parameter values. The highest level of accuracy achieved on the benchmark data set was 88.2% of the test programs correctly classified (a result also achieved under each of the four experimental scenarios). The best results were achieved in four instances, all where n>5 and L>5000. The highest level of classification accuracy achieved in the Student benchmark analysis was 88.5%, as shown in Table 2. This was achieved in one instance, for n=8 and L=3000. (Note that this level of performance was also achieved in two of the experiments conducted and was in fact exceeded under the final experimental scenario, as detailed in the following subsections.)

**Table 1.** Performance summary of the OSJava and corresponding identifier type data sets

|  | OSJava | OSSimple | OSClass | OSMethod | OSAll |
|---|---|---|---|---|---|
| **Mean classification accuracy** | 72.3% | 73.3% | 67.9% | 73.3% | 78.4% |
| **Median classification accuracy** | 76.5% | 76.5% | 70.6% | 76.5% | 82.4% |
| **Minimum classification accuracy** | 41.2% | 47.1% | 35.3% | 47.1% | 47.1% |
| **Maximum classification accuracy** | 88.2% | 88.2% | 88.2% | 88.2% | 88.2% |
| **Std. Deviation** | 13.3% | 12.7% | 13.9% | 12.1% | 13.6% |
| **Worse than OSJava** |  | 16 | 33 | 15 | 0 |
| **Better than OSJava** |  | 23 | 5 | 27 | 39 |
| **Same as OSJava** |  | 17 | 18 | 14 | 17 |

**Table 2.** Performance summary of the StudentJava and corresponding identifier type data sets

|  | StudentJava | StudentSimple | StudentClass | StudentMethod | StudentAll |
|---|---|---|---|---|---|
| **Mean classification accuracy** | 78.3% | 80.0% | 77.0% | 80.9% | 86.8% |
| **Median classification accuracy** | 76.9% | 80.8% | 76.9% | 80.8% | 88.5% |
| **Minimum classification accuracy** | 69.2% | 73.1% | 69.2% | 73.1% | 80.8% |
| **Maximum classification accuracy** | 88.5% | 88.5% | 84.6% | 88.5% | 92.3% |
| **Std. Deviation** | 4.4% | 4.3% | 3.6% | 4.6% | 3.5% |
| **Worse than StudentJava** |  | 5 | 18 | 5 | 0 |
| **Better than StudentJava** |  | 26 | 1 | 37 | 55 |
| **Same as StudentJava** |  | 25 | 37 | 14 | 1 |

## 6.1. Contribution of Simple Identifiers

The aim of the first experiment was to assess the degree to which simple identifiers defined by the programmer contributed to authorship identification. This category included all simple variable names that were defined within the program by the programmer. For example, variable names of type int, int[],long, char, boolean and so on were considered in this scenario.

All instances of these names were changed to a unique identifier comprised of a letter and a number followed by a different letter and the same number. (An example of one such unique identifier after disguising would be a15b15.) If the same variable name was used in more than one program these were changed to different unique identifiers, in order to eliminate the common byte level n-grams based on these variables (thus creating a conservative test). This task was performed on both benchmark data sets, OSJava and StudentJava, resulting in two data sets named OSSimple and StudentSimple. User-defined identifiers that remained unchanged for this experiment were all class variables and method names.

The results achieved for the OSsimple data set are shown, in summary, in Table 1. By comparing these results to those obtained from the analysis of the OSJava benchmark, it can be observed that accuracy remained the same in 17 of the 56 OSsimple n-L cases, it was improved in 23 cases and in the remaining 16 cases accuracy was worse. Using the one-tailed t-test to compare the mean levels of accuracy achieved with the OSJava and OSsimple data sets we found that the p-value was 0.0989, greater than our threshold of 0.05. Therefore the difference between the accuracy levels across the two data sets (at mean -1.06) is not statistically significant.

The summary results achieved from the StudentSimple data set are shown in Table 2. Comparing the results obtained from this analysis with the StudentJava benchmark results, we found that classification accuracy was unaffected in 25 of the 56 cases, in 26 cases we attained better results and in 5 cases classification accuracy decreased – thus again suggesting that classification accuracy improved with the disguising of simple identifiers. Using the one-tailed t-test of the difference between the StudentJava and StudentSimple mean accuracy (-1.72 as shown in Table 4), we found that the p-value was 0.0000. The difference between the two data sets is in this case was statistically significant.

The conclusion drawn from these rather mixed results is that, when analyzed using the SCAP method, simple variables in these programs do not play a significantly positive role in authorship attribution since accuracy did not deteriorate for either data set. The apparent improvement in accuracy achieved for many of the n, L combinations across both data sets but primarily in the StudentJava dataset appears to be due to the fact that many programmers use the same (or similar) names for simple variables. Some examples of the commonly used names encountered across different programmers were year, e, f, i, mid. The byte-level n-grams derived from these commonly used names were responsible for the initially incorrect classification of some programs in the benchmark analyses, particularly in the StudentJava data set. By making each user-defined simple variable unique in each program, we eliminated all these common n-grams across the different programmers, leading to an apparent improvement in overall classification accuracy.

Table 3. One-tailed t-test paired difference between the OSJava and corresponding identifier type data sets

|  | Mean | Std. Deviation | p-value |
|---|---|---|---|
| **OSJavaAccuracy - OSSimpleAccuracy** | -1.06 | 6.06 | 0.0989 |
| **OSJavaAccuracy - OSClassAccuracy** | 4.42 | 5.64 | 0.0000 |
| **OSJavaAccuracy - OSMethodAccuracy** | -1.06 | 6.05 | 0.0987 |
| **OSJavaAccuracy - OSAllAccuracy** | -6.07 | 5.48 | 0.0000 |

Table 4. One-tailed t-test paired difference between StudentJava and corresponding identifier type data sets

|  | Mean | Std. Deviation | p-value |
|---|---|---|---|
| **StudentJavaAccuracy - StudentSimpleAccuracy** | -1.72 | 2.94 | 0.0000 |
| **StudentJavaAccuracy - StudentClassAccuracy** | 1.31 | 2.24 | 0.0000 |
| **StudentJavaAccuracy - StudentMethodAccuracy** | -2.62 | 3.14 | 0.0000 |
| **StudentJavaAccuracy - StudentAllAccuracy** | -8.47 | 4.43 | 0.0000 |

### 6.2. Contribution of Class Identifiers

In the second set of experiments each name in the training and test programs from both data sets (OSJava and StudentJava) that pertained to the Class Identifiers category was changed to a unique identifier, following the same pattern as described above. Again, if the same class identifier was used in two different files, then it was changed to two different unique names. An example could be the class name 'Owner' that was used (possibly but not necessarily by a certain programmer) in two different programs. It would be changed to a123b123 in the first program and a34b34 in the second. The data sets thus derived are referred to as OSClass and StudentClass. (All the names that were either simple variables or method names remained unchanged.)

The accuracy results obtained in this experiment from the OSClass data set are shown in Table 1. Comparing these results with the OSJava benchmark results, it can be seen that in 33 of the 56 cases we had poorer author attribution performance, in 18 cases the same level of accuracy was achieved, and in 5 cases we achieved improved accuracy after identifier neutralization. The p-value obtained from the one-tailed t-test comparing the OSJava and OSClass mean accuracy values (being 4.42) was 0.0000, indicating that the disguising of class identifiers makes a statistically significant difference to classification accuracy.

The results obtained in this experiment from the StudentClass data set are shown in summary form in Table 2. When these results are compared with those obtained for the StudentJava benchmark experiment, it can be seen that in 18 of the 56 cases we had poorer attribution performance, in 37 cases the same level of accuracy was achieved, and in 1 case we achieved an improved result. The p-value obtained from the one-tailed t-test comparing the mean accuracy levels achieved for the StudentJava and StudentClass data sets (at 1.31) was 0.0000 (Table 4), illustrating that once again the difference in accuracies between the two data sets is statistically significant.

The results from these two experiments consistently indicate that class variables do appear to reflect program authorship because classification accuracy deteriorated in both data sets. While it is true that the mean difference between the StudentClass and StudentJava data sets is smaller than that found between the OSJava and OSClass sets, this is due in part to the fact that identifier naming in the StudentJava data set had been influenced to a degree by the instructor and the domain (being the same in most programs). In addition some of the programs had been plagiarized. Each of these factors would increase the likelihood of finding the same or similar class names across programs from different authors. Even when these inherently similar names were replaced by the unique identifiers (thus eliminating some of the common n-grams between programs from different authors), accuracy only improved in a single case. Thus, we can contend with evidence that class naming (i) is influential in reflecting authorship, and (ii) is reasonably robust to external influence and potentially to manipulation associated with the masking of plagiarism.

### 6.3. Contribution of Method Identifiers

This set of experiments was performed to evaluate the degree to which method names contributed to accurate authorship attribution. All method names defined within the programs in both the OSJava and StudentJava data sets were changed to unique identifiers. If the same method name appeared in more than one program, it was replaced by a different identifier in each case. All identifiers that were either simple variables or class variables were left unaffected. The resulting data sets are referred to as OSMethod and StudentMethod respectively.

The authorship attribution results for the OSMethod experiment are summarised in Table 1. Comparing these outcomes with the OSJava benchmark, the results are worse in 15 of the 56 cases, in 27 cases performance was improved and in 14 cases the same levels of accuracy were achieved. The next step was to perform a one-tailed t-test to evaluate the significance of the difference between the accuracies achieved using the OSJava and OSMethod programs (at a value of -1.06). The p-value of this test was found to be 0.0987, higher than our threshold for p of 0.05. This shows that the difference between the accuracies in these data sets is not significant.

The results achieved from the analysis of the StudentMethod data set are given in Table 2. Comparing these levels of accuracy against those obtained for the StudentJava benchmark, the results are worse in 5 of the 56 cases, in 37 cases performance was improved and in 14 cases the same levels of accuracy were achieved. The p-value obtained by comparing the StudentJava and StudentMethod mean accuracy values (at -2.62) was 0.0000 (see Table 4). This shows that the SCAP analysis of the StudentMethod data set, with its method names disguised, achieved statistically better classification accuracy than achieved through the analysis of the StudentJava data set.

The improvement in accuracy evident mainly in the StudentJava data set, despite the disguising of method names, is again explained in part by the fact that the unique identifiers that replaced the user-defined method names eliminated some of the common n-grams that were derived from coincidentally common or similar method names used by different programmers (and that negatively affected the level of correct classification achieved in the benchmark tests). Examples of such names included getInteger, setString, init set. The degree of improvement observed after identifier neutralization is greater in the instructor-influenced single domain StudentJava data set because the number of common method names used by different programmers is higher in this set. The conclusion drawn from these results is that method names defined by the user in these Java programs do not play a significantly positive role in authorship attribution using the SCAP method since accuracy did not deteriorate in either data set.

### 6.4. Contribution of All User-defined Identifiers

One further experiment was conducted to assess the impact of neutralizing *all* names, belonging to all three categories considered above. In this experiment all identifiers including simple variables, method names and class variables defined by the programmer within each program in the OSJava and StudentJava data sets were replaced by unique identifiers, including those that appeared in more than one program (irrespective of authorship). The purpose of this experiment was to assess the extent of influence that all names used within a program had on authorship attribution. The data sets derived are referred to as OSAll and StudentAll correspondingly.

The summary results obtained for the OSAll data set are reported in Table 1. By comparing these results to those obtained from the analysis of the OSJava benchmark, it can be observed that accuracy remained the same in 17 of the 56 cases and was in fact improved in the other 39 cases. This indicates that, in this case, the names defined by the users did not contribute positively to authorship attribution. The p-value obtained in comparing the OSJava and OSAll mean accuracies using a paired-sample t-test was 0.0000. This shows that the levels of classification accuracy obtained from the two analyses are significantly different.

Similarly the summary results from the StudentAll data set are shown in Table 2. A comparison of the StudentJava and StudentAll classification performance reveals that accuracy was the same in 1 of the 56 cases and was improved in the other 55. This again provides evidence that the user-defined names in these programs did not contribute positively to authorship attribution. The t-test p-value obtained in comparing the two mean accuracies was 0.0000, suggesting that the levels of accuracy achieved in the analysis of the two data sets are significantly different.

The improvement in classification accuracy obtained after the disguising of identifiers is highest in this last experiment, for both data sets – not unexpected given the results obtained in the three preceding tests. Examination of the analyses revealed that, as for the prior experiments, this improvement can be explained by the fact that programs written by different programmers contained the same or similar names (for example value and val, or fragment, fragmentation, fragmentname and fragments). The byte-level n-grams derived from these commonly used names were responsible for the originally incorrect classification of some programs in the benchmark analyses. By making each user-defined identifier unique in each program we eliminated all these common n-grams across the different programmers, thus improving overall classification accuracy when compared to the two benchmark sets.

## 7. CONCLUSIONS

We have performed a number of experiments in order to assess the impact of different Java identifier types on source code authorship attribution, using the Source Code Author Profile approach. In these experiments, programs from two different Java data sets with different characteristics were used. The first data set contained open source code and programs that were 'domain independent' since all programs from each author that were placed in the training set were from a different project than the programs placed in the test set. Hence, the programs in this data set did not share common characteristics. In contrast, the second data set was formed by programs written during an introductory Java course, the consequence being that naming in these programs was influenced by the instructor and that some program samples had been plagiarized. In addition most programs in this data set belonged to the same application domain. As a result the programs in this data set shared several common characteristics and identifiers.

In each experiment one category of identifiers was neutralized, in order to provide a means of measuring the difference between classification accuracy with and without the certain type of identifier available. The results of these experiments (presented in Tables 1 to 4) have shown the following for the data sets assessed here:

- Simple variables and method names defined by the programmer do not seem to positively influence classification accuracy since accuracy did not deteriorate in either data set. In fact in some cases accuracy could be improved if these names were neutralized before the SCAP analysis. This is due to the fact that programmers have been shown to use the same or similar names for simple variables and method names. This conclusion applied to both Java data sets considered here but to a lesser extent for the OSJava data set, where the programs were from a different application domain for each programmer.
- Class and Object naming does positively influence authorship classification accuracy, an outcome evident for programs in both data sets, since accuracy deteriorated in both data sets after disguising class and object names.
- Accuracy classification is improved by neutralizing all user-defined identifiers in both data sets. This conclusion has also been reached in the study of Frantzeskou et al [4] using the OSJava dataset. In the current study this result has been confirmed to hold in a different data set; namely the StudentJava.

At the outset of this study we asked the following questions: Do Java identifiers contribute to correct authorship identification? Is it possible to hide the provenance of some Java program by identifier renaming? The results of our analyses suggest that the answer to the first question is a partial 'Yes', Java *class* identifiers do contribute to correct authorship identification. The answer to the second question appears to be 'No' – it is not possible to hide the provenance of some Java program by identifier renaming. In fact, globally renaming all identifiers – neutralizing them – enabled us to actually improve our authorship classification accuracy over the benchmark levels achieved with identifiers intact.

One of the implications of our work is that future Java authorship identification systems that are intended to explain why it is claimed that a piece of code is written by a particular author should concentrate on the class identifiers in analysing and assigning authorship. More broadly, identifier neutralization could be used as a means of improving accuracy in Java authorship identification cases. In contexts in which identifiers might be named in 'standard' ways the masking of identifiers (perhaps apart from class names) should be performed before authorship analysis in undertaken. A related implication is that the plagiarism of source code may be more difficult to mask than perpetrators believe, in that the global disguising of identifiers – a common method employed by those who plagiarise – may in fact make the original authorship more evident, rather than less.

We acknowledge that this is just one set of experiments, and that further work should be done with larger and different samples. We ourselves have undertaken further research with Lisp code [4]. Future work could include research on other specific programming languages, in order to check in detail whether our findings are language- or data set-specific. Also of interest would be research investigating whether novice developers do indeed adopt naming conventions characteristic of their textbooks or instructors, and the impact of such adoption on classification accuracy.